\documentclass[11pt]{article}
\usepackage{amssymb}
\usepackage{graphicx}% to include figures
\DeclareFontFamily{U}{rsfs}{\skewchar\font"7F}
\DeclareFontShape{U}{rsfs}{m}{n}{
	<-6> rsfs5
	<6-8> rsfs7
	<8-> rsfs10
	}{}
\DeclareMathAlphabet{\mathscr}{U}{rsfs}{m}{n}

\newcommand{\bra}{\langle}
\newcommand{\ket}{\rangle}
\newcommand{\prob}{\mbox{Prob}}

\begin{document} % ----------------------------------------
\baselineskip 6.0mm % mark
\hfill % \today % date
\begin{flushright}
\begin{minipage}{60mm}
{\small
quant-ph/0703118
\\
1st version, March 14, 2007
% 3rd version June 11, 2008
% revised thoroughly, June 9, 2015
\\
submitted to Quanta, June 9, 2015
\\
revised, July 13, 2015
}
\end{minipage}
\end{flushright}
\vspace*{12mm}
\begin{center}
% title
{\bf \Large
Complementarity and the Nature of Uncertainty Relations 
\vspace{2mm} \\
in Einstein-Bohr Recoiling Slit Experiment}
\vspace{4mm}

Shogo Tanimura\footnote{E-mail: tanimura[AT]is.nagoya-u.ac.jp}\footnote{Published in Quanta Vol 4, No 1, 1-9 (2015). 
http://quanta.ws/ojs/index.php/quanta/article/view/35
DOI:10.12743/quanta.v4i1.35. 
}
\vspace{4mm}

{\it
Department of Complex Systems Science,
Graduate School of Information Science,
\\
Nagoya University,
Nagoya 464-8601, Japan
}

\vspace{10mm}

Abstract
\vspace{3mm}

\begin{minipage}{110mm}
\baselineskip 5.0mm % mark
A model of the Einstein-Bohr double-slit experiment 
is formulated in a fully quantum theoretical setting.
In this model,
the state and dynamics of a movable wall that has the double slits in it,
as well as the state of a particle incoming to the double slits,
are described by quantum mechanics.
Using this model,
we analyzed complementarity between 
exhibiting the interference pattern and distinguishing the particle path.
Comparing 
% the two types of uncertainty relations,
the Kennard-Robertson type and the Ozawa-type uncertainty relations,
we conclude that 
the uncertainty relation involved in the double-slit experiment is 
not the Ozawa-type uncertainty relation
but the Kennard-type uncertainty relation of the position and the momentum
of the double-slit wall.
A possible experiment to test the complementarity relation is suggested.
It is also argued that
various phenomena which occur at the interface 
of a quantum system and a classical system,
including distinguishability, interference, decoherence, 
quantum eraser, and weak value,
can be understood as aspects of entanglement.
\end{minipage}
\vspace*{14mm}

\begin{minipage}{130mm}
\baselineskip 5.0mm % mark
Keywords: double-slit experiment, uncertainty relation, 
Kennard-Robertson inequality, Ozawa inequality,
entanglement, complementarity.
% interference, visibility, distinguishability
\end{minipage}

\end{center}

\vspace*{3pt}

\newpage
\baselineskip 5.3mm % mark
\section{Introduction}
The uncertainty relation is one of the best known subjects 
which manifest the peculiar nature of the microscopic world.
Although many people have been discussing it for a long 
time~\cite{Heisenberg}-\cite{Caves1985},
% time~\cite{Heisenberg, Neumann, Kennard, Robertson, Braginskii, Caves1980, Yuen, Caves1985},
some confusion 
about the formulation and the implication of the uncertainty relation 
remained.
Recently, Ozawa~\cite{Ozawa1988, Maddox}
settled down the controversy about the uncertainty relation
and he established a new inequality~\cite{Ozawa2003}, which represents
a quantitative relation between 
measurement error and disturbance caused by measurement.

According to his 
formulation~\cite{Ozawa2003}-\cite{Ozawa2004},
% formulation~\cite{Ozawa2003, Ozawa1984, Ozawa2004},
a measurement process is described as an interaction process 
of an observed object and an observing apparatus.
Suppose that the object has observables $ \hat{A} $ and $ \hat{B} $.
The apparatus has a meter observable $ \hat{M} $,
which is designed to point the value of $ \hat{A} $.
The whole system is initialized at the time $ t = 0 $
and the measurement is made at a later time $ t $.
The readout of the meter is represented by the operator
$ \hat{M} (t) 
:= e^{i \hat{H} t / \hbar} \hat{M} e^{-i \hat{H} t / \hbar} $ 
and the true value of the object observable is $ \hat{A} (0) := \hat{A} $.
Their difference $ \hat{N} := \hat{M}(t) - \hat{A}(0) $ is called a noise operator.
The change in $ \hat{B} $,
$ \hat{D} := \hat{B}(t) - \hat{B}(0) $, is called a disturbance operator.
These expectation values
\begin{eqnarray}
&&	\varepsilon (\hat{A}) 
	:= \sqrt{ 
	\Big\bra ( \hat{M}(t) - \hat{A}(0) )^2 \Big\ket } \, ,
	\label{error} \\
&&	\eta (\hat{B}) 
	:= \sqrt{ 
	\Big\bra ( \hat{B}(t) - \hat{B}(0) )^2 \Big\ket } \, ,
	\label{disturbance} \\
&&	\sigma (\hat{A}) 
	:= \sqrt{ 
	\Big\bra ( \hat{A} - \bra \hat{A} \ket )^2 \Big\ket } \, ,
%	\qquad
	\\
&&	\sigma (\hat{B}) 
	:= \sqrt{ 
	\Big\bra ( \hat{B} - \bra \hat{B} \ket )^2 \Big\ket }
%	:= \sqrt{ \bra \hat{B}^2 \ket - \bra \hat{B} \ket^2 } 
\end{eqnarray}
are defined with respect to the initial state of the whole system.
The quantity 
$ \varepsilon (\hat{A}) $ 
is the error involved in the measurement of $ \hat{A} $.
The quantity 
$ \eta (\hat{B}) $ 
is the disturbance in $ \hat{B} $ caused by the measurement.
The quantity 
$ \sigma(\hat{A}) $ 
is the standard deviation of $ \hat{A} $ in the initial state.
It is reasonable to call $ \sigma(\hat{A}) $ the statistical fluctuation.

A naive expression of the uncertainty relation
\begin{equation}
	\varepsilon (\hat{q}) \, \eta (\hat{p}) \, \gtrsim \, h
	\qquad
	\mbox{(wrong)}
	\label{Heisenberg}
\end{equation}
for the position $ \hat{q} $ and the momentum $ \hat{p} $ of a particle
is sometimes attributed to Heisenberg.
Originally, Heisenberg~\cite{Heisenberg} examined
a thought experiment of a gamma-ray microscope
for investigating limit of accuracy of measurement 
on a microscopic particle
and concluded the relation (\ref{Heisenberg}).
He stated that the microscope is
an example of the destruction of the knowledge of particle's momentum 
by an apparatus determining its position~\cite{Heisenberg book}.
It should be mentioned that
Heisenberg himself did not give the rigorous definitions of
error, disturbance, and statistical fluctuation.
He did not distinguish these notions, either.
Therefore, it seems hard 
to shift the responsibility of the inequality (\ref{Heisenberg}) onto Heisenberg.

Von Neumann~\cite{Neumann} formulated 
a model of a measurement process
and proved the inequality
$ \varepsilon (\hat{q}) \, \eta (\hat{p}) \, \ge \, \frac{1}{2} \hbar $,
but his proof apparently depends on the specific model.
Kennard~\cite{Kennard} gave a mathematical proof of the inequality
\begin{equation}
	\sigma (\hat{q}) \, \sigma (\hat{p}) \, \ge \, \frac{1}{2} \hbar
	\label{Kennard}
\end{equation}
in a model-independent manner.
Robertson~\cite{Robertson} proved a more general relation
% into the form
\begin{equation}
	\sigma (\hat{A}) \, \sigma (\hat{B}) \, \ge \, \frac{1}{2} 
	\Big| \bra [ \hat{A}, \hat{B} ] \ket \Big|
	\label{Robertson}
\end{equation}
for arbitrary observables $ \hat{A} $ and $ \hat{B} $.
Considering their implications,
we call the inequalities (\ref{Kennard}) and (\ref{Robertson})
{\it the standard-deviation uncertainty relations}
or {\it the relations of fluctuations intrinsic to a quantum state}.
It should be noted that 
the Kennard-Robertson relation (\ref{Kennard}), (\ref{Robertson})
has nothing to do with the measurement apparatus.
% The Kennard-Robertson inequalities have been regarded
% as a mathematically rigorous formulation of the uncertainty relation
% but actually they do not represent the relation of 
% measurement error and disturbance.
% physical implication that Heisenberg originally aimed to formulate.

Ozawa~\cite{Ozawa2003} formulated a general scheme of measurement
and introduced the rigorous definitions of
error and disturbance, (\ref{error}) and (\ref{disturbance}).
Using them he proved the inequality
\begin{equation}
	\varepsilon (\hat{A}) \, \eta (\hat{B}) 
	+ \varepsilon (\hat{A}) \, \sigma (\hat{B}) 
	+ \sigma (\hat{A}) \, \eta (\hat{B}) 
	\, \ge \, \frac{1}{2} 
%	\hbar
	\Big| \bra [ \hat{A}, \hat{B} ] \ket \Big|.
	\label{Ozawa}
\end{equation}
Moreover, he constructed concrete models~\cite{Ozawa2003, Ozawa2004}
that yield $ \varepsilon (\hat{q}) = 0 $ and $ \eta (\hat{p}) = $ finite.
Hence, the Ozawa inequality (\ref{Ozawa}) is correct
while the naive uncertainty relation (\ref{Heisenberg}) does not hold in general. 
It seems suitable to call the inequality (\ref{Ozawa})
{\it the error-disturbance uncertainty relation}
or {\it the relation of indetermination involved in a measurement process}.
% His argument settled down the dispute on the physical implication of
% the uncertainty relation.
Later, Branciard \cite{Branciard2013} proved a tighter inequality.
% than the Ozawa inequality (\ref{Ozawa}).

% On the other hand, % mark
The interference effect of matter wave, 
or the particle-wave duality of matter,
is another well-known peculiarity of quantum mechanics.
When a beam of particles is emitted toward a wall that has double slits on it,
we observe an interference pattern on a screen behind the wall.
Quantum mechanics tells that,
if we put some device 
to detect which slit each particle has passed,
the interference pattern disappears.
It is impossible to distinguish the path of each particle
without smearing the interference pattern.
This kind of incompatibility 
of exhibitions of the particle property and the wave property
is called {\it complementarity} by Bohr.
% simultaneously.
% mark

% Originally, 
Einstein proposed, in his debate with Bohr, a thought experiment
in which 
the wall with the double slits is allowed to move to detect a collision of the particle.
He argued that 
% if one knew both the position and the momentum 
% of the double-slit wall simultaneously,
one could detect on which slit each particle recoils
without destroying the interference pattern.
Bohr~\cite{Bohr} answered that,
because of the uncertainty relation,
one cannot determine simultaneously the position and the momentum of the wall
and hence distinguishing the particle path 
and viewing the interference pattern
are incompatible.
%as a consequence of the uncertainty relation of
%the position and the momentum of the wall.
% in this way;
% Because of the uncertainty relation,
% the simultaneous measurements of the position and the momentum of the wall
% is impossible.
% If the uncertainty relation holds,
% distinguishability of the particle path 
% and visibility of the interference pattern remains incompatible.
Similar explanations can be found 
in some textbooks~\cite{Messiah, Feynman}, too.
Recently, Miron et al.~\cite{Miron2015} made
experimental realization of a movable double-slit system.

Although the double-slit experiment is regarded as 
a pedagogical subject from the viewpoint of modern physics,
it remains unclear what kind of uncertainty relation is involved there. 
Thus we propose a question;
{\it Which type of the uncertainty relation,
the Kennard inequality (\ref{Kennard})
or the Ozawa inequality (\ref{Ozawa}),
prevents the simultaneous measurements of the interference and the path?}
% It seems an interesting question to ask
% which of the Kennard inequality or the Ozawa inequality 
% is relevant to
% the double-slit experiment.
This is the question we study in this paper.

In this paper
we formulate a model of the double-slit experiment 
in genuine quantum theoretical terms
and analyze % quantitatively % relation
the complementarity 
between distinguishing the particle path
and viewing the interference pattern.
Our conclusion is that the complementarity
involves the Kennard-type uncertainty relation,
which is the property intrinsic to the quantum state 
of the double-slit wall.
% We also obtain a quantitative relation between the distinguishability and the visibility.
We propose an experiment
to test this distinguishability-visibility relation.

\section{Model and its analysis}
\begin{figure}[tb]
\begin{center}
%\scalebox{0.60}{
\includegraphics{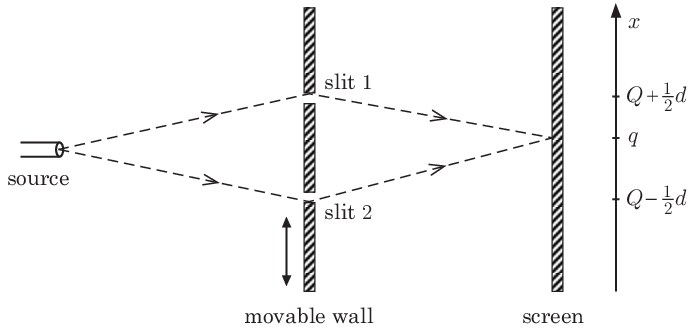}
%}
\end{center}
\vspace{-2mm}
\caption{\label{double-slit}The Young interferometer
in the Einstein-Bohr scheme.}
\end{figure}

Here we shall formulate a model of the Young interferometer
of the form proposed in the Einstein-Bohr discussion~\cite{Bohr}.
As shown in Fig.~\ref{double-slit}, 
a particle is emitted from the source,
flies through the double slits on the wall, 
and arrives at the screen behind the wall.
We call each slit as
% Each slit is specified as 
slit~1 and slit~2, respectively.
They are separated by a distance $ d $.
The coordinate axis, which we call the $ x $-axis, 
is taken to be parallel to the wall and the screen.
The wall is movable along the $ x $-axis
and the screen is fixed.
The coordinate and the momentum 
of the particle are denoted as $ (q,p) $.
% On the other hand,
Similarly,
the coordinate and the momentum of the double-slit wall
are denoted as $ (Q,P) $.
The $ x $-coordinate of slit~1 is $ Q + \frac{d}{2} $
while
the $ x $-coordinate of slit~2 is $ Q - \frac{d}{2} $.
A position eigenstate of the whole system is 
$ | q \ket \otimes | Q \ket = | q, Q \ket $.
The initial state of the whole system is assumed to be
\begin{equation}
	| \mbox{initial} \ket
	=
	| \phi \ket \otimes | \xi \ket,
	\label{initial state}
\end{equation}
which is a composite of a particle state $ | \phi \ket $ 
with a wall state $ | \xi \ket $.
The emitted particle obeys the free-particle Hamiltonian.
Its time-evolution operator is 
\begin{equation}
	\hat{U} ( t ) = \exp \Big( - \frac{i}{\hbar} \frac{\hat{p}^2}{2m} t \Big).
\end{equation}
We assume that the particle reaches the slits on the movable wall at the time $ \tau $.

The function of the slits are described by two projection operators,
$ \hat{S}_1 $ and $ \hat{S}_2 $.
If the particle arrives at slit 1, its state becomes
$ \hat{S}_1 | \phi \ket $.
If the particle arrives at slit 2, its state becomes
$ \hat{S}_2 | \phi \ket $.
They satisfy 
$ \hat{S}_{\alpha}^\dagger = \hat{S}_{\alpha} $,
$ \hat{S}_{\alpha}^2 = \hat{S}_{\alpha} $,
$ \hat{S}_1 + \hat{S}_2 = 1 $ and $ \hat{S}_1 \hat{S}_2 = \hat{S}_2 \hat{S}_1 = 0 $.
It is assumed that 
the particle gives a momentum $ +k $ to the wall
when the particle hits slit 1.
On the other hand,
the particle gives a momentum $ -k $ to the wall
when the particle hits slit 2.
This interaction is described by the Hamiltonian
\begin{equation}
	\hat{V} 
	=
	\hat{S}_1 F ( \hat{q} - \hat{Q} )
	- \hat{S}_2 F ( \hat{q} - \hat{Q} ).
\end{equation}
Here $ F $ is a constant force.
The evolution operator for an infinitesimal time interval $ \Delta t $ is
\begin{equation}
%	e^{ - \frac{i}{\hbar} \hat{V} \Delta t }
	e^{ - i \hat{V} \Delta t / \hbar}
	=
	e^{
	- i \hat{S}_1 k ( \hat{q} - \hat{Q} )  / \hbar
	\, + \, i \hat{S}_2 k ( \hat{q} - \hat{Q} )  / \hbar}.
%	\frac{i}{\hbar} \hat{S}_1 k (\hat{Q}-\hat{q}) 
%	- \frac{i}{\hbar} \hat{S}_2 k (\hat{Q}-\hat{q}) }.
\end{equation}
Here $ k := F \Delta t $ is the impact (momentum transfer).
The operator $ e^{ik\hat{Q} / \hbar} $ shifts the wall momentum 
as $ e^{ik\hat{Q} / \hbar} | P \ket = | P + k \ket $
while the operator $ e^{-ik\hat{q} / \hbar} $ shifts the particle momentum 
as $ e^{-ik\hat{q} / \hbar} | p \ket = | p - k \ket $.
% The operator $ e^{-ik\hat{q} / \hbar} \otimes e^{ik\hat{Q} / \hbar} $ can be regarded
Hence, after the momentum exchange, the state of the whole system becomes
\begin{equation}
	| \mbox{behind slits} \ket
	=
	e^{-ik\hat{q} / \hbar} \hat{S}_1 \hat{U}(\tau) | \phi \ket \otimes 
	e^{ ik\hat{Q} / \hbar} | \xi  \ket 
	\, + \,
	e^{ ik\hat{q} / \hbar} \hat{S}_2 \hat{U}(\tau) | \phi \ket \otimes 
	e^{-ik\hat{Q} / \hbar} | \xi  \ket.
	\label{split state}
\end{equation}
The particle arrives at the screen at the time $ \tau + \tau' $.
Then the state of the whole system becomes
\begin{eqnarray}
	| \mbox{final} \ket
	&=&
	\hat{U}(\tau') e^{-ik\hat{q} / \hbar} \hat{S}_1 \hat{U}(\tau) | \phi \ket \otimes 
	e^{ ik\hat{Q} / \hbar} | \xi  \ket 
	\, + \,
	\hat{U}(\tau') e^{ ik\hat{q} / \hbar} \hat{S}_2 \hat{U}(\tau) | \phi \ket \otimes 
	e^{-ik\hat{Q} / \hbar} | \xi  \ket
\nonumber \\
	&=&
	| \phi_1 \ket \otimes 
	e^{ ik\hat{Q} / \hbar} | \xi  \ket 
	\, + \,
	| \phi_2 \ket \otimes 
	e^{-ik\hat{Q} / \hbar} | \xi  \ket.
	\label{reflected state}
\end{eqnarray}
Here we put 
$ \hat{U}(\tau') e^{-ik\hat{q} / \hbar} \hat{S}_{1} \hat{U}(\tau) | \phi \ket = | \phi_{1} \ket $,
$ \hat{U}(\tau') e^{ ik\hat{q} / \hbar} \hat{S}_{2} \hat{U}(\tau) | \phi \ket = | \phi_{2} \ket $.
% $ \hat{U}(\tau') e^{-ik\hat{q} / \hbar} \hat{S}_{\alpha} \hat{U}(\tau) | \phi \ket = | \phi_{\alpha} \ket $.
Then the probability for finding the particle at the position $ q $ on the screen
is proportional to
\begin{eqnarray}
	\prob (q) 
	& \propto &
	\int \Big| \bra q, Q \, | \mbox{final} \ket \Big|^2
	dQ
\nonumber \\
	&=&
	\big| \phi_1 (q) \big|^2 +
	\big| \phi_2 (q) \big|^2 +
	2 \, \mbox{Re} \Big\{
	\phi_1^* (q) \phi_2 (q) \,
	\bra \xi | e^{-2ik\hat{Q} / \hbar} | \xi \ket \Big\}
\nonumber \\
	&=&
	\big| \phi_1 (q) \big|^2 +
	\big| \phi_2 (q) \big|^2 +
	2 \, {\mathscr V} \, \mbox{Re} \Big\{
	e^{i \alpha} \, \phi_1^* (q) \phi_2 (q) 
	\Big\}.
	\label{interference}
\end{eqnarray}
The last term describes an interference of the two waves
$ \phi_1 (q) = \bra q | \phi_1 \ket $ and 
$ \phi_2 (q) = \bra q | \phi_1 \ket $.
The nonnegative real number $ {\mathscr V} $ and the phase $ e^{i \alpha} $
are defined by
\begin{equation}
	{\mathscr V} e^{i \alpha} =
	\bra \xi | e^{-2ik\hat{Q} / \hbar} | \xi \ket.
	\label{visibility}
\end{equation}
The contrast of the interference fringe is proportional to $ {\mathscr V} $,
which is called the visibility of the interference
and takes its value in the range $ 0 \le {\mathscr V} \le 1 $.
The wavefunction of the movable wall is denoted as
$ \xi(Q) = \bra Q | \xi \ket $
and its Fourier transform is
\begin{equation}
	\tilde{\xi} (P) 
	= \frac{1}{\sqrt{2 \pi \hbar}} \int e^{-iPQ/\hbar} \, \xi (Q) dQ.
	\label{Fourier2}
\end{equation}
In terms of the wall wavefunctions, the visibility is written as
\begin{equation}
	{\mathscr V} 
	= 
	\left|
	\int
	\xi^* (Q) \, e^{-2ikQ / \hbar} \, \xi (Q) dQ
	\right|
	= 
	\left|
	\int
	\tilde{\xi}^* (P-k) \, \tilde{\xi} (P+k) dP
	\right|.
	\label{V}
\end{equation}

After observing the position of the particle on the screen,
we measure the momentum of the wall to detect the path of the particle.
The conditional probability distribution of the momentum $ P $ is calculated from
(\ref{reflected state}) as
\begin{eqnarray}
	\prob (P|q)
	\: \propto \:
	\Big| \bra q, P \, | \mbox{final} \ket \Big|^2
	=
	\left|
		\phi_1 (q) \, \tilde{\xi} (P-k)
		+
		\phi_2 (q) \, \tilde{\xi} (P+k)
	\right|^2.
	\label{probability for P}
\end{eqnarray}
If the support of the initial wavefunction
$ | \tilde{\xi} (P)| $ is contained within the range
$ P_0 - k < P < P_0 + k $ for some $ P_0 $,
then from the measured value of $ P $
we can tell the slit which the particle passed.
Namely,
if the measured momentum is in the range
$ P_0 < P < P_0 + 2k $,
we can say that the particle hit slit~1.
On the other hand,
if the measured momentum is in the range
$ P_0 - 2k < P < P_0 $,
we can say that the particle hit slit~2.
However, 
if the support of $ | \tilde{\xi} (P)| $ is contained within
$ P_0 - k < P < P_0 + k $,
the overlap integral in (\ref{V}) vanishes
and hence the interference fringes fade away completely.

Contrarily,
if the width of the support of $ \tilde{\xi} (P) $ is larger than $ 2k $,
the visibility (\ref{V}) can be nonzero.
However, at that time, 
the probability (\ref{probability for P}) can have a 
nonzero interference term
and hence we cannot distinguish the particle path certainly.

We summarize the above argument symbolically as
\begin{eqnarray*}
	\mbox{Visible interference}
	& \Leftrightarrow &
	{\mathscr V} \ne 0
\\
	& \Rightarrow &
	\mbox{supp} \, | \tilde{\xi} (P-k) | \cap
	\mbox{supp} \, | \tilde{\xi} (P+k) | \; \;
	\mbox{has nonzero measure}.
\\
	& \Leftrightarrow &
	\mbox{The path of the particle cannot be distinguished completely}
	\\
	&& \quad \mbox{by measuring the momentum of the wall.}
\end{eqnarray*}
In the above inference,
the second arrow $ ( \Rightarrow ) $
cannot be replaced with the necessary and sufficient sign $ ( \Leftrightarrow ) $.
For example, if we take the wavefunction
\begin{equation}
	\tilde{\xi} (P) =
	\left\{
		\begin{array}{ll}
			a \, \sin \left( \frac{2 \pi P}{2k} \right)
			& ( 0 \le P \le 2k )
		\\
			b \, \sin \left( \frac{4 \pi P}{2k} \right)
			& ( -2k \le P \le 0 )
		\\	
			0
			& ( \mbox{otherwise} ),
		\end{array}
	\right.
\end{equation}
then the supports of 
$ | \tilde{\xi} (P-k)| $ and $ | \tilde{\xi} (P+k)| $ 
have an overlap with nonzero measure
but the integral $ {\mathscr V} $ vanishes.

\section{Uncertainty relation}
Now we discuss 
what kind of uncertainty relation is involved in the double-slit experiment.
Let us consider the expression (\ref{V}) for the visibility,
\begin{equation}
	{\mathscr V} (k)
	= 
	\left|
	\int
	e^{-2ikQ / \hbar} \, | \xi (Q) |^2 \, dQ
	\right|.
	\label{probability for q via Q}
\end{equation}
Suppose that the probability distribution $ | \xi(Q) |^2 $
has an effective width $ \Delta Q $.
(A rigorous definition of $ \Delta Q $ is not necessary for the following argument.)
Since (\ref{probability for q via Q}) is an oscillatory integral,
to get a considerably large visibility we need to have
\begin{equation}
	2 k \, \Delta Q / \hbar \: \lesssim \: 2 \pi.
	\label{Delta Q}
\end{equation}
On the other hand, as discussed above,
to distinguish the slit which the particle passes,
the initial momentum distribution of the double-slit wall 
should be contained within the range
\begin{equation}
	\Delta P < 2k.
	\label{Delta P}
\end{equation}
Hence, 
to observe a clear interference 
and to distinguish the path of the particle simultaneously,
we need to have 
\begin{equation}
	\Delta Q \, \Delta P \: \lesssim \: 2 \pi \hbar.
	\label{Delta Q Delta P <}
\end{equation}
As a contraposition,
the uncertainty relation
\begin{equation}
	\Delta Q \, \Delta P \: \gtrsim \: 2 \pi \hbar
	\label{Delta Q Delta P >}
\end{equation}
implies that
exhibiting a clear interference pattern and 
detecting the particle path
cannot be accomplished simultaneously.
This uncertainty relation (\ref{Delta Q Delta P >})
is a property of the initial state of the double-slit wall
but it is not a relation 
between error and disturbance caused by measurement.
Hence, we conclude that
{\it the obstruction 
against the simultaneous realization of interference and path detection
is the Kennard-type uncertainty relation 
which is intrinsic to the quantum state of the double-slit wall.}
This is the main claim of this paper.
It is to be noted that this is a conclusion of an analysis of the specific model.
We do not have to take it as a universally valid statement.

% old mark
% On the other hand,
It is a long-standing issue
% there was a puzzle asking
whether the position-momentum uncertainty relation does imply or not 
the particle-wave complementarity.
Storey {\it et al}.~\cite{Storey1994, Storey1995} argued that
the position-momentum uncertainty relation is responsible for destroying
the interference pattern.
Englert {\it et al}.~\cite{Scully1991, Scully1995} took an opposite position
and argued that 
the {\it which-way information} is responsible for destroying
the interference pattern.
% and that the disturbance in the momentum is not necessarily responsible.
D{\"u}rr, Nonn, and Rempe~\cite{Durr, DurrPRL}
performed experiments to prove that 
what destroys the interference pattern is the which-way information.
Hence, Englert's argument seems to be correct.
However, 
momentum disturbance is still necessary for the change of the interference pattern
in the which-way experiment,
and Englert did not answer to this point.

What we discussed in this paper is a question asking
{\it which kind of the position-momentum uncertainty relation
% the Kennard type (\ref{Kennard}) or the Ozawa type (\ref{Ozawa}),
is responsible for destroying the interference pattern
when we try to detect the path of the particle 
by measuring the momentum of the movable slit-wall}.
Our answer is that
{\it the Kennard-type uncertainty relation is responsible
in the context of our model.}

\section{Suggestions for experiments}
\begin{figure}[bt]
\begin{center}
% \scalebox{0.60}{
\includegraphics{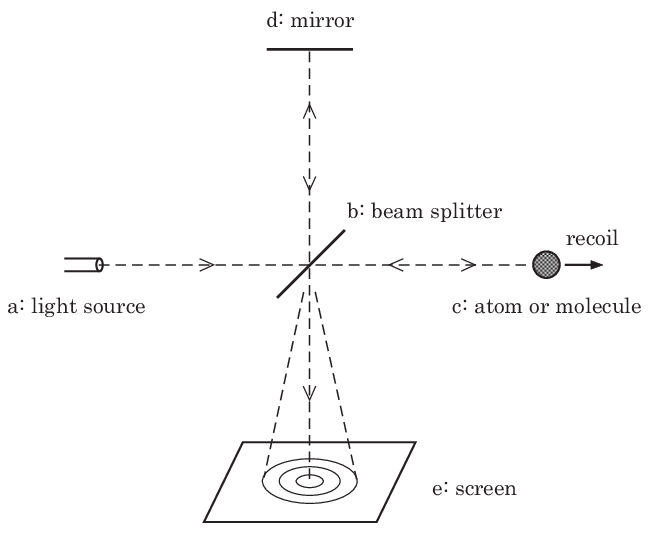}
%}
\end{center}
\vspace{-2mm}
\caption{\label{Michelson}The Michelson interferometer.}
\end{figure}

Here we would like to suggest an experimental scheme to test
the visibility relation (\ref{V}). % old mark
Our scheme uses the Michelson interferometer as illustrated in Fig.~\ref{Michelson}.
A photon is emitted from a light source (a)
and is split by a beam splitter (b) into two directions.
At the end (c) of one direction,
an atom or a molecule is placed.
The incident photon is scattered by the atom and the atom recoils.
At the other end (d) a mirror is fixed.
The two paths merge at the beam splitter (b)
and the photon reaches the fixed screen (e).
There we observe an interference pattern 
by accumulating photons.
On the other hand, 
by measuring the velocity of the atom, we can infer the path of the photon;
If the atom recoils out, we know that the photon took the path (c).
If the atom remains stationary, we know that the photon took the path (d).

If the initial wavefunction of the atom is strongly localized, 
one will observe a clear interference pattern 
but fail to determine the velocity of the atom precisely.
If the initial wavefunction of the atom has a larger spatial spread, 
one can determine the velocity of the atom with a smaller error
but the interference pattern will become feebler.
Thus the initial state of the atom defines 
the visibility of the interference fringe as (\ref{V}).
We may put a Bose-Einstein condensate (BEC) of atoms at the place (c)
instead of a single atom
since control and observation of the BEC are more feasible than a single atom.

We can estimate the velocity of the recoiling atom.
It is assumed that
the photon has wave length $ \lambda $
and the target atom has mass $ M $.
Then the photon momentum is $ p = h/\lambda $
and the impact given to the atom is $ k = 2p $.
The velocity of the recoil atom is
\begin{equation}
	v = \frac{k}{M} = \frac{2h}{M \lambda}.
\end{equation}
Assume that 
the photon wave length is $ \lambda = 0.5 \times 10^{-6} $ m
and that we use a mercury atom as a reflector.
Then the recoil velocity is
$ v = 7.9 \times 10^{-3} \, {\rm m} \cdot {\rm s}^{-1} $.
If we use a BEC of $ 10^4 $ sodium atoms, the velocity is
$ v = 6.9 \times 10^{-6} \, {\rm m} \cdot {\rm s}^{-1} $.
On the other hand,
% Furthermore, 
the argument around Eq. (\ref{Delta Q}) implies that
the size of the spread of the wavefunction of the target should be smaller than
\begin{equation}
	\Delta Q 
	\sim \frac{\pi \hbar}{k} 
	=    \frac{1}{4} \lambda
	\label{distinguishable delta Q}
\end{equation}
for exhibiting a clear interference pattern.
% distinguishing the path which the photon takes.

In the above argument we proposed a use of the Michelson interferometer.
Other interference experiments,
like the Hanbury-Brown-Twiss correlation~\cite{HBT1, HBT2}
or the interference of photons from two light sources,
which has been demonstrated by Mandel {\it et al.}~\cite{two-photon1, two-photon2}, 
can also be modified to experiments
which demonstrate the tradeoff
% a dilemma 
between interference and distinguishability.
It is also to be noted that
Plau {\it et al.}~\cite{Pfau} and Chapman {\it et al.}~\cite{Chapman} 
had demonstrated
that a change of the momentum distribution of an atom
by photon emission or by photon scattering causes a change 
of spatial coherence of the atom.
They had confirmed the Kennard-type uncertainty relation.

% mark
After completing this manuscript, we have learned that
a group of Miron~\cite{Miron2015}
successfully observed interference pattern of electrons emitted from two atoms.
They observed also disappearance of the interference fringes 
when the atoms were not fixed
and the atom emitting an electron recoiled by the momentum transfer.

\section{Entanglement as a root of measurement effects}
Various phenomena including 
distinguishability, interference and decoherence, 
quantum eraser, and weak value,
which occur at the interface between quantum systems and classical systems,
are related to measurement and entanglement.
Before closing our discussion we would like to explain that
these phenomena can be understood on the same footing.

Assume that an object system is in a superposed state
$ c_1 | \phi_1 \ket +c_2 | \phi_2 \ket $
and an apparatus is in a state $ | \xi \ket $.
The vectors $ | \phi_1 \ket $, $ | \phi_2 \ket $ are unit vectors,
but not necessarily orthogonal.
The initial state of the whole system is the tensor product state
\begin{equation}
	\Big( c_1 | \phi_1 \ket + c_2 | \phi_2 \ket \Big) \otimes | \xi \ket.
%	\label{initial state}
\end{equation}
Interaction between the object and the apparatus entangles 
the two systems to the state
\begin{equation}
	| \Psi \ket
	=
	\hat{U} \Big( c_1 | \phi_1 \ket + c_2 | \phi_2 \ket \Big) \otimes | \xi \ket
	=
	c_1 | \phi_1 \ket \otimes | \xi_1 \ket +
	c_2 | \phi_2 \ket \otimes | \xi_2 \ket .
	\label{measured state}
\end{equation}
It is assumed that 
the vectors $ | \xi_1 \ket $, $ | \xi_2 \ket $ are unit vectors,
but not necessarily orthogonal.
The object has an observable $ \hat{A} $
and the apparatus has an observable $ \hat{M} $.
The spectral decompositions of these observables are
\begin{equation}
	\hat{A} = \sum_a a \, \hat{P}_A (a),
	\qquad
	\hat{M} = \sum_m m \, \hat{P}_M (m).
\end{equation}
In the followings, we explain that
various aspects of measurement can be understood
as properties of the entangled state (\ref{measured state}).

i) Which-way distinguishability: 
In the context of which-way experiment,
the apparatus is designed for distinguishing
the object states $ | \phi_1 \ket $ and $ | \phi_2 \ket $.
After the interaction, the probability for reading out the value $ m $
from the meter observable $ \hat{M} $ is given by
\begin{eqnarray}
	\prob ( \hat{M} = m) 
&=& 
	\bra \Psi | \hat{P}_M (m) | \Psi \ket
	\nonumber \\
&=& 
	|c_1|^2 \bra \xi_1 | \hat{P}_M (m) | \xi_1 \ket 
	+|c_2|^2 \bra \xi_2 | \hat{P}_M (m) | \xi_2 \ket
	\nonumber \\ && \,
	+ \, 2 \, \mbox{Re} \, \big\{
	c_1^* c_2 \, \bra \phi_1 | \phi_2 \ket \,
	\bra \xi_1 | \hat{P}_M (m) | \xi_2 \ket 
	\big\}. 
	\label{prob m}
\end{eqnarray}
If the third term 
$ \bra \phi_1 | \phi_2 \ket \, \bra \xi_1 | \hat{P}_M (m) | \xi_2 \ket $
vanishes,
and if either
$ \bra \xi_1 | \hat{P}_M (m) | \xi_1 \ket $ or
$ \bra \xi_2 | \hat{P}_M (m) | \xi_2 \ket $ vanishes,
we successfully distinguish the states 
$ | \phi_1 \ket $ and $ | \phi_2 \ket $
by reading the value $ m $.
Oppositely,
if the both terms
$ \bra \xi_1 | \hat{P}_M (m) | \xi_1 \ket $ and 
$ \bra \xi_2 | \hat{P}_M (m) | \xi_2 \ket $
are nonzero, we have error for distinguishing the states 
$ | \phi_1 \ket $ and $ | \phi_2 \ket $.
If the third term 
$ \bra \phi_1 | \phi_2 \ket \, \bra \xi_1 | \hat{P}_M (m) | \xi_2 \ket $
is nonzero, the distinguishing measurement fails, too.
When the meter states $ | \xi_1 \ket $ and $ | \xi_2 \ket $ are orthogonal,
the meter becomes optimal for distinguishing
the states $ | \phi_1 \ket $ and $ | \phi_2 \ket $.

ii) Interference visibility and decoherence:
If we measure the quantity $ \hat{A} $ directly,
the probability for obtaining the value $ a $ of $ \hat{A} $ is
\begin{eqnarray}
	\prob ( \hat{A} = a) 
&=&
	\bra \Psi | \hat{P}_A (a) | \Psi \ket
	\nonumber \\
&=&
	|c_1|^2 \bra \phi_1 | \hat{P}_A (a) | \phi_1 \ket 
	+|c_2|^2 \bra \phi_2 | \hat{P}_A (a) | \phi_2 \ket
	\nonumber \\ && \,
	+ \, 2 \, \mbox{Re} \, \big\{
	c_1^* c_2 \, \bra \phi_1 | \hat{P}_A (a) | \phi_2 \ket \,
	\bra \xi_1 | \xi_2 \ket 
	\big\}. 
	\label{prob a}
\end{eqnarray}
The interference effect is characterized by the coefficient
$ c_1^* c_2 $, which depends on the phases of $ c_1 $ and $ c_2 $.
When the matrix element
$ \bra \phi_1 | \hat{P}_A (a) | \phi_2 \ket $
is nonzero, the interference effect is observed.
But the contrast of interference fringe is reduced by the factor
$ \bra \xi_1 | \xi_2 \ket $.
When the meter states $ | \xi_1 \ket $ and $ | \xi_2 \ket $ is orthogonal,
the which-way distinguishing measurement is optimized
but the interference effect is completely lost.

iii) Quantum eraser: 
The above argument tells that 
the distinguishability is maximized but 
the interference effect is lost
when $ \bra \xi_1 | \xi_2 \ket = 0 $.
However, even when $ \bra \xi_1 | \xi_2 \ket = 0 $,
by a joint measurement of $ \hat{A} $ and $ \hat{M} $
the interference effect is recovered.
The joint probability for observing the values
$ \hat{A} = a $ and $ \hat{M} = m $ is
\begin{eqnarray}
	\prob ( \hat{A} = a, \hat{M} = m) 
&=&
	|c_1|^2 \bra \phi_1 | \hat{P}_A (a) | \phi_1 \ket 
	\bra \xi_1 | \hat{P}_M (m) | \xi_2 \ket 
	\nonumber \\ && \,
	+ \, |c_2|^2 \bra \phi_2 | \hat{P}_A (a) | \phi_2 \ket
	\bra \xi_1 | \hat{P}_M (m) | \xi_2 \ket 
	\nonumber \\ && \,
	+ \, 2 \, \mbox{Re} \, \big\{
	c_1^* c_2 \, \bra \phi_1 | \hat{P}_A (a) | \phi_2 \ket \,
	\bra \xi_1 | \hat{P}_M (m) | \xi_2 \ket 
	\big\}. 
	\label{prob a and m}
\end{eqnarray}
Note that
$ \bra \xi_1 | \hat{P}_M (m) | \xi_2 \ket $ can be nonzero
even when $ \bra \xi_1 | \xi_2 \ket = 0 $.
Thus the interference effect is observed by the joint measurement of
$ \hat{A} $ and $ \hat{M} $
although the interference effect is not observed by the measurement of
$ \hat{A} $ alone.
However, the which-way distinguishability is reduced
as $ \bra \xi_1 | \hat{P}_M (m) | \xi_2 \ket \ne 0 $.
In a sense, 
the which-way information is erased for reviving the interference fringe,
and hence this effect is named quantum eraser.
The conditional probability 
for obtaining the value $ a $ of $ \hat{A} $ 
under the restriction $ \hat{M} = m $ is calculated as
\begin{eqnarray}
	\prob ( \hat{A} = a | \hat{M} = m) 
&=&
	\frac{\prob ( \hat{A}=a, \hat{M}=m )}{\prob ( \hat{M}=m)}.
	\label{prob a/m}
\end{eqnarray}

iv) Weak value:
The weak probability is 
the conditional probability 
for obtaining the meter value $ m $ of $ \hat{M} $ 
under the selection of object value $ \hat{A} = a $,
\begin{eqnarray}
	\prob ( \hat{M} = m | \hat{A} = a ) 
&=&
	\frac{\prob ( \hat{A}=a, \hat{M}=m )}{\prob ( \hat{A}=a)}.
	\label{prob m/a}
\end{eqnarray}
When the probability $ \prob ( \hat{A}=a ) $ is small,
that means the measurement disturbance is small,
the joint probability $ \prob ( \hat{M} = m | \hat{A} = a ) $
becomes large.
Thus a kind of enhancement or amplification of the meter value can occur.
This is mechanism of the so-called weak measurement.

% mark
Before summarizing our discussion,
we give a brief overview of the developments in this area.
Englert~\cite{Englert} formulated a qualitative relation between
distinguishability and interference visibility.
D{\"u}rr and Rempe~\cite{Durr2000:AmJPhys}
gave a general proof of the Englert formula
using the noncommutativity of observables.
D{\"u}rr {\it et al.}~\cite{DurrPRL} tested this formula by experiment.
Hosoya {\it et al.}~\cite{Hosoya} investigated the complementarity
of which-way distinguishability and interference
from the viewpoint of entanglement.

Scully and Dr{\"u}hl~\cite{Scully1982} introduced the idea of quantum eraser.
Experimental realization of quantum eraser have been achieved repeatedly,
for example, 
as demonstrated by Hillmer and Kwiat~\cite{Hillmer2007}.

Aharonov, Albert and Vaidman~\cite{Aharonov1988} introduced the idea of weak value
as an extended notion of the values of physical observables.
The weak value is a value of meter conditioned by a postselected object value.
It was first pointed out by Tamate {\it et al.}~\cite{Tamate2009}
that the structure of the weak value (\ref{prob m/a})
is dual to the structure of the quantum eraser (\ref{prob a/m}).

\section{Summary}
In this paper we analyzed
the double-slit scheme that 
exhibits the interference pattern on the screen and
distinguishes the path of a particle by measuring momentum of the wall.
This double-slit setting has an entangled state of the particle and the wall.
Finally, we concluded that the Kennard-type uncertainty relation
of fluctuations of position and momentum of the wall is a reason of
the complementarity 
%of distinguishability and interference
in the double-slit experiment.
Similar analysis has been done by Qureshi and Vathsan~\cite{Qureshi2013}.
It is also to be noted 
Busch and Shilladay presented a study 
of the complementarity in Mach-Zehnder interferometry~\cite{Busch2006}.

\section*{Acknowledgements}
The author thanks 
Toshihiro Iwai, Mikio Nakahara, and Kazuya Yuasa
% T. Iwai, M. Nakahara, and K. Yuasa 
for their kind comments 
for this work.
The first version of the manuscript was submitted to e-print arXiv in March 2007
and labeled quant-ph/0703118.
The title is changed from the first version.
This work is financially supported 
by Japan Society for the Promotion of Science, % (JSPS), 
Grant Nos.~15540277, 17540372, and 26400417.

% \newpage

\end{document}